**Interaction of 3D mesostructures composed of Pd-Ni alloy nanowires with low-temperature oxygen plasma**

G.K. Strukova[1]*, G.V. Strukov[1], S.V. Egorov[1], A.A. Rossolenko[1], D.V. Matveyev[1], V.S. Stolyarov[2,3], S.A. Vitkalov[4]

1 Institute of Solid State Physics, Russian Academy of Sciences, 142432, Chernogolovka, Russia
2 Moscow Institute of Physics and Technology, 141700, Dolgoprudny, Russia
3 Kazan Federal University, 420008, Kazan, Russia
4 Physics Department, City College of City University of New York, NY 10031, USA

* corresponding autor-Galina Strukova, e-mail: strukova@issp.ac.ru

**Abstract**

It has been found that volumetric mesoporous structures composed of Pd-Ni nanowires exhibit active interaction with low temperature non-equilibrium oxygen plasma, which results in a light red incandescence of the samples. The incandescent process lasts for about 10 min. After plasma-chemical treatment, the diffraction pattern revealed the presence of a nanocrystalline nickel (II) oxide phase on a surface of the samples. Magnetic measurements showed that a temperature dependence of the magnetic susceptibility as well as a magnetic hysteresis loop of the sample exposed in the plasma is typical for metals and similar to the dependence for the original Pd-Ni sample. The composite mesoporous structures produced by the plasma-chemical treatment have a conducting Pd-Ni frame covered with a nickel (II) oxide (semiconductor) and are of interest as a promising material for electronic devices.

## 1. Introduction

In the search of new nanostructured materials with a large surface area, suitable for practical applications, researchers concentrate their attention toward the hierarchical 3D structures consisting of metallic or metal oxide nanowires. Previously, we reported on an effective electrochemical method for the fabrication of mesoporous hierarchical structures composed of metallic nanowires [1-4]. Very recently, 3D structures of nanowires $Pt_3Co$ with an enhanced electro-catalytic activity were obtained by a hydrothermal method [5]. Also, 3D hierarchical ZnO nanowire structures that are of interest for a variety of applications have been fabricated by means of electrochemical anodization of zinc foil [6]. Hierarchical composite nanostructures containing conductive and robust 3D skeleton composed of metallic nanowires, which are coated with a metal-oxide semiconducting layer, are promising multipurpose material for applications. The standard oxygen plasma treatment of the mesostructures built of metallic nanowires looks attractive as an effective technological route for the fabrication of composite metal-semiconductor materials. In recent work Ni nanowires, derived from alumina membrane and placed on Si substrate, have been treated by means of the oxygen plasma. However no nickel oxide have been found in the treated samples [7]. The oxygen plasma treatment of the hierarchical 3D nanostructures has not been investigated so far. This paper presents investigations of 3D mesoporous structures composed of Pd-Ni nanowires exposed to low temperature non-equilibrium oxygen plasma. The study indicates a presence of Ni (II) oxide in the treated samples.

## 2. Experimental

The method of the fabrication of metallic nanostructured mesostructures is presented in [2,3]. The studied samples were treated with oxygen plasma in a cylindrical quartz reactor (diameter 95 mm, length 300 mm) with a cylindrical radio-frequency electrode placed over its entire length and grounded end plugs. The generated power achieved 100 W. Oxygen was blown through the cylinder at a rate from 4.5 to 5 ml/min., the oxygen pressure being from 0.7 to 0.8 mbar. The samples were studied using a Leica M165 C optical stereo microscope and SUPRA 50VP, JEOL 7001F, JEM-2100 scanning electron microscopes. Measurements of the dynamic magnetic susceptibility of the samples were carried out at a frequency of 100 kHz. VSM-155 vibrating coil magnetometer was used in measurements of hysteresis loops.

## 3. Results

Fig.1 presents convex-concave mesoscopic structures composed of Pd-Ni nanowires ("seashells") that were grown by the described techniques [2,3].

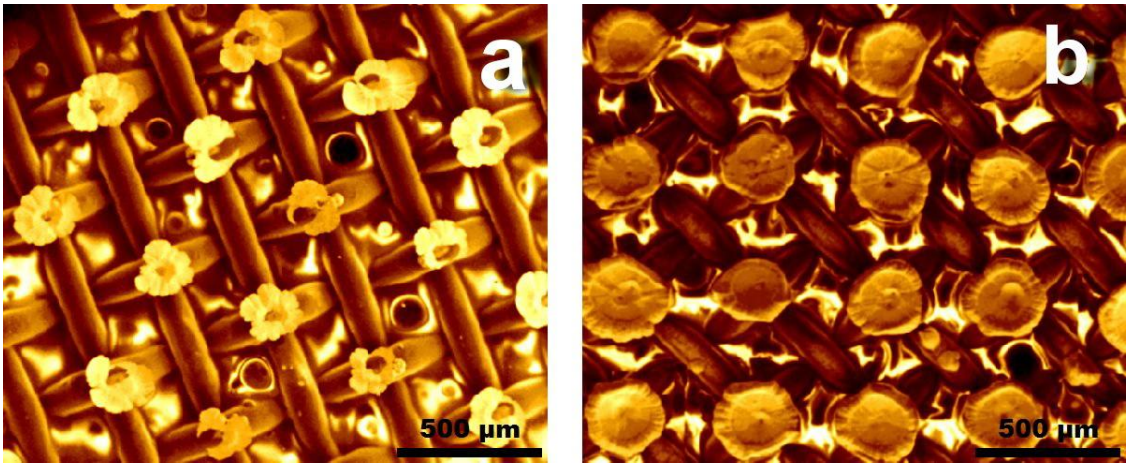

**Fig.1.** Array of convex-concave Pd-Ni mesostructures grown on net: "laced" (a) and "massive" (b).

The "massive" and "laced" samples have an identical composition of Pd-Ni, typically 15-35 at.% Pd. The "massive" samples are obtained by a prolonged electrodeposition. The characteristic pattern on the inner surface of the sample presented in Fig.2a.b is a manifestation of the volumetric frame formed due to a self-organization of growing nanowires [3].

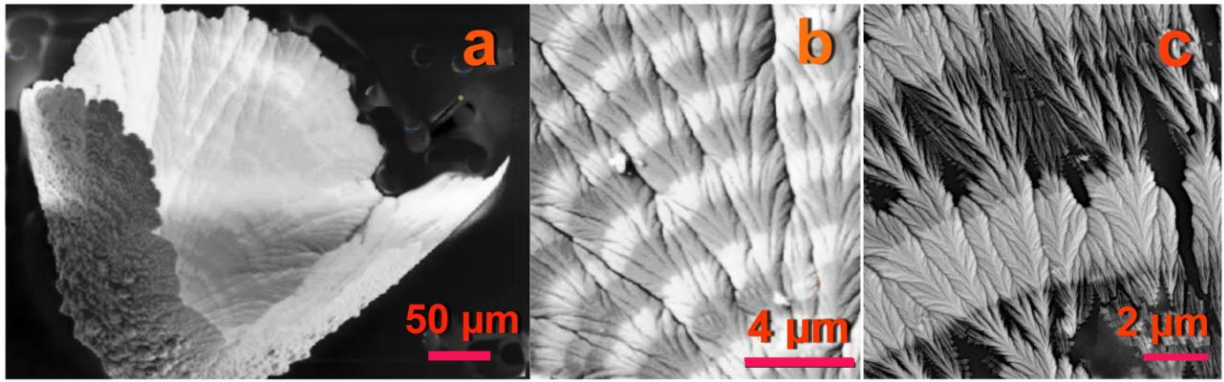

**Fig.2.** Pd-Ni "seashell": SEM images of a "laced" sample (a), its inner surface (b), and bundles of nanowires under the thin shell covering the "seashell" wall (c).

The seashell wall is covered by a thin outer layer [3]. A removal of the outer layer of Pd-Ni "seashell" by a chemical polishing exposes a porous network of nanowires shown in Fig.3.

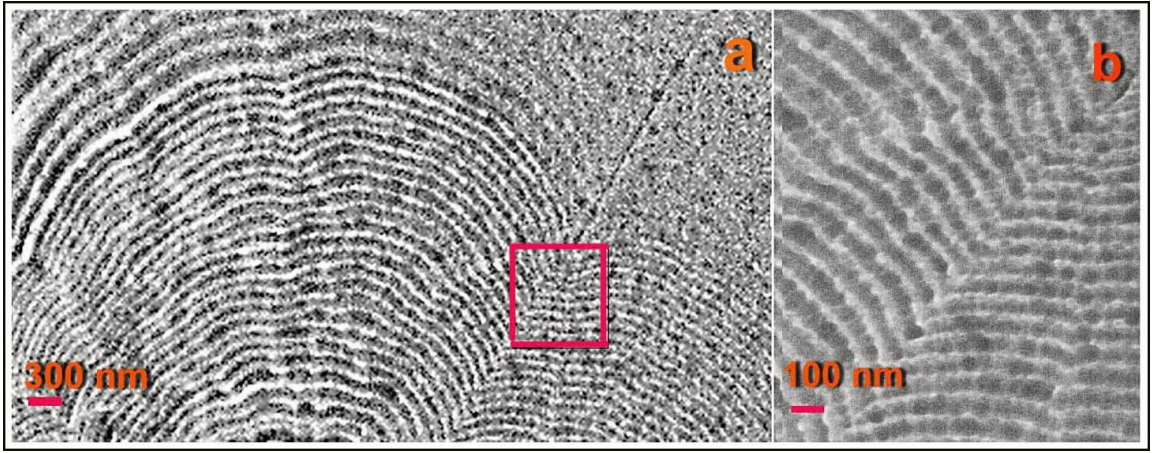

**Fig.3.** SEM images of Pd-Ni "seashell" internal structure demonstrating network of nanowires.

SEM images presented in Fig.3 allow estimating the porosity of the seashell wall as the ratio of the area occupied by pores (dark areas) to the entire cross-sectional area. The estimations yield a high porosity (≈60-65%) of the studied samples.

The obtained samples of Pd-Ni mesostructures were then treated by low-temperature oxygen plasma in the plasma-chemical reactor following the procedure described in the experimental section. Contacting with the oxygen plasma, the samples react immediately with a light red incandescence. The incandescence reaction persists as long as the plasma generator is switched on (typically about 10 min.). Fig. 4 shows the images of the Pd-Ni "seashells" before and after the oxygen plasma treatment obtained in an optical stereo-microscope.

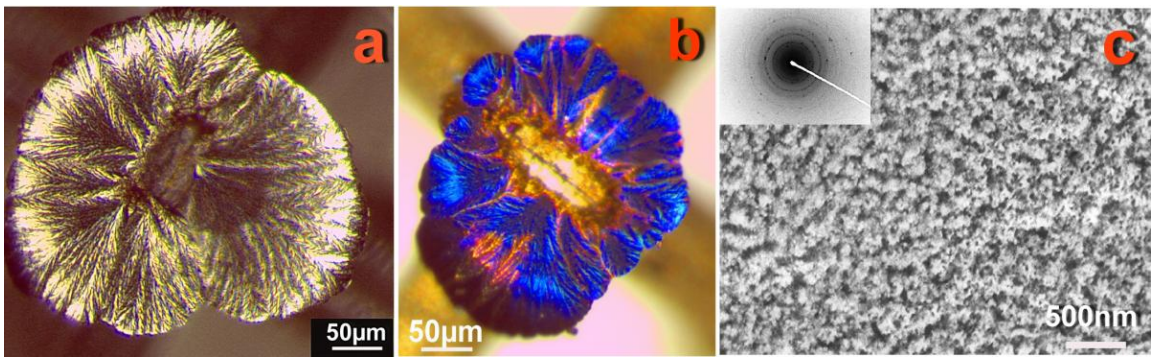

**Fig. 4**. Optical images of Pd-Ni samples: before (a) and after oxygen plasma treatment for 10 min (b). SEM image of the surface of the plasma-treated sample (c). The insert shows the corresponding electron diffraction pattern indicating the nickel (II) oxide phase.

After the oxygen plasma treatment the inner surface of the sample retains the characteristic pattern indicating the preserved original frame located under the surface layer. The surface of the plasma-treated samples demonstrates a dark blue color which is inherent to the oxide nickel (II) films having the properties of p-type semiconductors [8]. The EDX analysis of the surface of the plasma treated samples show a considerable increase of the oxygen content (more then 10%) and a decreased amount of the nickel and palladium compared to the untreated samples. Furthermore Fig.4c shows that the surface consists of crystallites less than 50 nm in size.  Shown in the insert electron diffraction pattern indicates that the crystallites are formed by the nickel (II) oxide phase with the cell parameters 0,417 nm  corresponding to this phase. In contrast, shown in Fig.5a the temperature dependence of the magnetic susceptibility of the "seashells" treated in oxygen plasma is nearly identical to the dependence of the pristine Pd-Ni samples. The magnetic hysteresis loops of both pristine Pd-Ni and plasma-treated samples show no appreciable difference indicating a weak influence on the oxygen plasma treatment on the bulk properties of the studied composite structures (Fig.5 b). The presented observations signal that a thin NiO layer is formed on the metallic frame. Thus, the oxygen plasma treatment transforms the mesostructures built of Pd-Ni nanowires into a composite containing the metallic Pd-Ni frame, which is covered with nano-crystalline semiconducting nickel oxide.

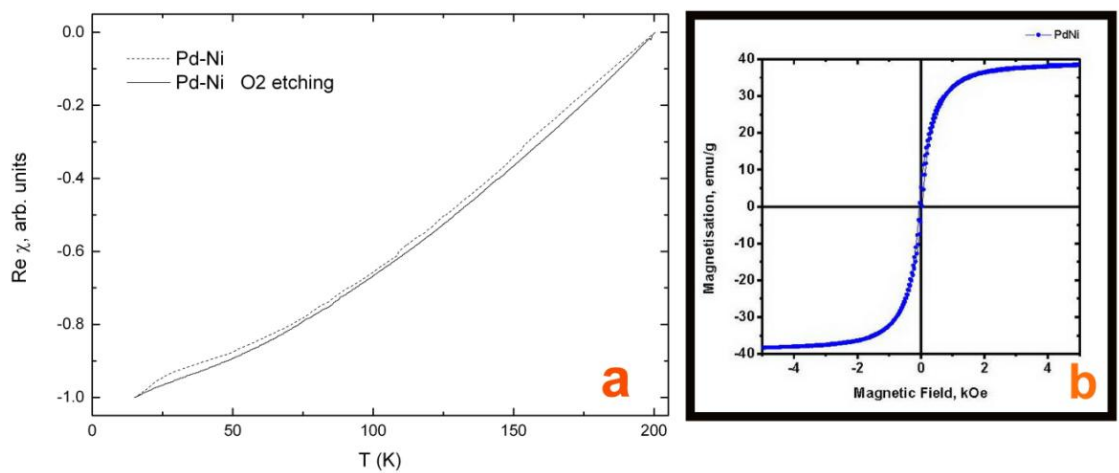

**Fig.5**. Temperature dependences of magnetic susceptibility (a) and magnetic hysteresis loops (b) of original and plasma-treated Pd-Ni sample.

Also, both  as the magnetic hysteresis loop of pristine Pd-Ni sample and the same of plasma-treated its showed no appreciable differences apparently due to a weak influence on it of a thin layer of nickel oxide (Fig.5 b). These facts are the evidence that a thin NiO layer is formed on the metallic frame. Thus, the oxygen plasma treatment changes the mesostructure composed of Pd-Ni nanowires into a composite in which the metallic Pd-Ni frame is covered with nanocrystalline nickel oxide.

## 4. Discussion

The instantaneous incandescence reaction of Pd-Ni mesostructures with the oxygen plasma occurs likely due to the effect of an increased internal energy of samples composed of nanowires.

SEM studies of identical samples have shown [3] that the Pd-Ni "seashells" are composed of nanowires with various diameters from 20 to 100 nm. The nanowires are shaped as beads, that greatly increases their total surface area. As a result of the self-organization, the nanowires assemble into conic bundles and arrange into a volumetric 3D network with numerous pores. In such nanostructured porous constructions most of atoms sit near the surface increasing significantly the free energy and decreasing, therefore, the stability of the structures. The internal energy is further increased due to contributions of a large number of defects and the amorphous content between the nanocrystals that is typical for the studied structures in question [4]. Fig. 6 shows the results of TEM studies of the nanostructure identical Pd-Ni samples [4].

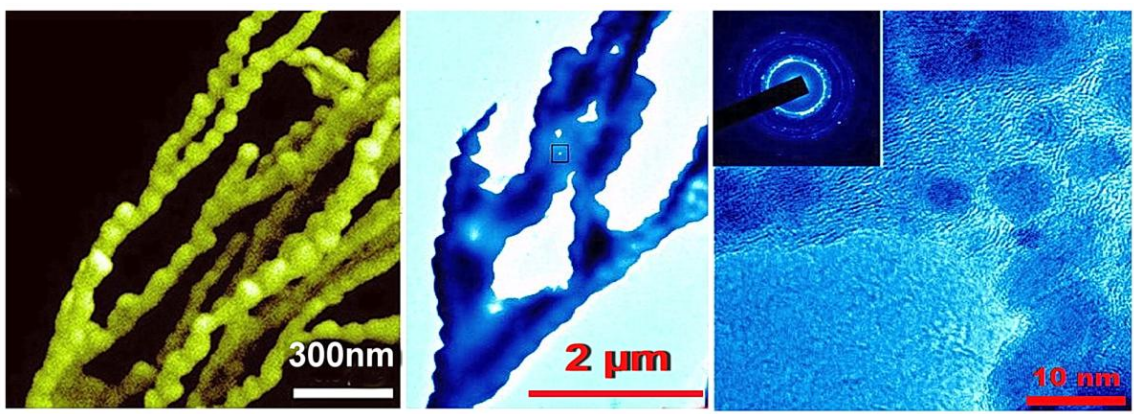

**Fig. 6.** Pd-Ni "seashell": TEM images of the nanowire branch (two pictures on the left); magnified high-resolution image of the area selected in the middle picture (right); electron diffraction image of selected area (insert) [4].

Both bundles of nanowires consisting Pd-Ni "seashell" and individual nanowires were separated by a treatment in an ultrasonic bath. The nanowires look like V-branched "beads". The high-resolution image shown in Fig.6 demonstrates clearly that the nanowires consist of 4-15 nm Pd-Ni crystallites immersed in Pd-Ni amorphous matrix. The electron diffraction pattern shown in the insert to Fig.6 reveals two fcc phases with cell parameters of 0,363 nm and 0,384 nm indicating a heterogeneous content of the Pd-Ni phase composition.

The increased internal energy due to a large surface area and a large number of defects, together with the presence of an amorphous phase are the apparent sources of the high rate of the observed incandescent process and the subsequent surface oxidation. We note that a similar treatment of Ni nanowires of 200 nm in diameter and 30 μm in length with fcc crystalline structure in oxygen plasma did not produce the nickel surface oxidation [7]. Clearly, that an amorphous-nanocrystalline structure of nanowires and the dense porosity of studied structures facilitate significantly the heterogeneous reaction of the nickel atoms with the oxygen plasma. The surface atoms located in the amorphous content (a metastable state) are less bounded in a comparison with the ones in the crystal lattice and, thus, are more active to react with the oxygen. Due to the large surface area (high porosity and small diameter of the nanowires), a large amount of the metal atoms located near the surface of the nanowires are available for the contact with the oxygen plasma. These atoms react instantaneously with energy-active particles of the plasma ("hot" electrons and positive ions) and chemically-active particles (free oxygen atoms in an excited state) exciting the incandescence effect.

It is reasonable to assume that the oxidation reaction of nickel atoms, Ni + O = NiO, proceeds only in the thin near-surface layer. The fact is that the energy of the reaction is only 0.66 of NiO sublimation energy. Therefore, the nickel oxide film, formed in this reaction, is hardly removable by the plasma. The oxide film covers the entire nanowire surface isolating the bulk metal atoms from the reaction with the plasma. The Pd-Ni nanowire oxidation reaction thereby is expected to be limited in time. For this reason, the long incandescence of the sample cannot be explained only by the exothermal reaction of the nickel atoms with the free oxygen atoms of the plasma. This phenomenon requires additional studies.

**4.1. Conclusion**

Thus, the oxygen plasma treatment of the mesostructures composed of Pd-Ni metallic nanowires produces a composite mesostructure "NiO on Pd-Ni frame". Electronic properties of these hierarchical mesostructures with large surface area are of significant interest for further research. For instance, authors of a recent work [9] observed peculiarities in local electron transport of p-type nickel oxide nanostructures grown on substrates that may be used in electronic devices.

**5. Outlook**

The recently proposed method [1-4] fabricates hierarchical 3D mesostructures composed of nanowires or nanoclusters of various normal, magnetic, and superconducting metals and alloys. The synergy of the method with the plasma-chemical treatment using nitrogen, air, hydrogen sulphide, ammonia and other gases opens new possibilities for creating a wide variety of nanostructured hierarchical systems of "semiconductor-on-metal frame". Research of the physical properties of these novel mesostructures with a large surface area may be valuable for electronic and optoelectronic elements.

## 6. Acknowledgements

The authors thank Artem F. Shevchun, Yuriy P. Kabanov for the measurements of magnetic properties, Andrey A. Mazilkin for identifying phase NiO on the sample surface, and Valery V. Ryazanov, Leonid P. Mezhov-Deglin, Igor I. Khodos for a useful discussions.